\documentclass[conference]{IEEEtran}

\usepackage[cmex10]{amsmath}
\usepackage{amssymb}
\usepackage{graphicx}
\usepackage[capitalise]{cleveref}

\ifCLASSOPTIONcompsoc	\usepackage[caption=false,font=normalsize,labelfont=sf,textfont=sf]{subfig}
\else \usepackage[caption=false,font=footnotesize]{subfig}
\fi

\ifCLASSOPTIONcompsoc \usepackage[nocompress]{cite}
\else                 \usepackage{cite}
\fi

\usepackage{tikz}
\usetikzlibrary{fit,shapes.misc}
\newcommand\marktopleft[1]{%
    \tikz[overlay,remember picture] 
        \node (marker-#1-a) at (0,1.5ex) {};%
}
\newcommand\markbottomright[1]{%
    \tikz[overlay,remember picture] 
        \node (marker-#1-b) at (0,0) {};%
    \tikz[overlay,remember picture,thick,dashed,inner sep=3pt]
        \node[draw,rounded rectangle,fit=(marker-#1-a.center) (marker-#1-b.center)] {};%
}

\begin{document}

\title{Wi-Fi Rate Adaptation using a Simple Deep Reinforcement Learning Approach}

\author{
    \IEEEauthorblockN{
        Rúben Queirós,
        Eduardo Nuno Almeida,
        Helder Fontes,
        José Ruela,
        Rui Campos
    }
    \IEEEauthorblockA{
        INESC TEC and Faculdade de Engenharia, Universidade do Porto, Portugal \\
        \{ruben.m.queiros, eduardo.n.almeida, helder.m.fontes, jose.ruela, rui.l.campos\}@inesctec.pt
    }
}
\maketitle

\begin{abstract}
The increasing complexity of recent Wi-Fi amendments is making optimal Rate Adaptation (RA) a challenge. The use of classic algorithms or heuristic models to address RA is becoming unfeasible due to the large combination of configuration parameters along with the variability of the wireless channel. Machine Learning-based solutions have been proposed in the state of art, to deal with this complexity. However, they typically use complex models and their implementation in real scenarios is difficult.
 
We propose a simple Deep Reinforcement Learning approach for the automatic RA in Wi-Fi networks, named Data-driven Algorithm for Rate Adaptation (DARA). DARA is standard-compliant. It dynamically adjusts the Wi-Fi Modulation and Coding Scheme (MCS) solely based on the observation of the Signal-to-Noise Ratio (SNR) of the received frames at the transmitter. Our simulation results show that DARA achieves up to 15\% higher throughput when compared with Minstrel High Throughput (HT) and equals the performance of the Ideal Wi-Fi RA algorithm.
\end{abstract}


\section{Introduction}

The Wi-Fi standard has been evolving with recent amendments, such as IEEE 802.11n (Wi-Fi 4), IEEE 802.11ac (Wi-Fi 5), and more recently IEEE 802.11ax (Wi-Fi 6/6E). New configuration parameters have been added  to both the physical  (PHY)  and  medium  access  control  (MAC)  layers. However, the high variability of the radio channel, allied to the channel asymmetry, makes the optimal configuration of these parameters challenging. The Modulation and Coding Scheme (MCS) is one example very closely related to the quality of communication links and their stability \cite{ra_eval}. MCS optimization improves the efficiency of IEEE 802.11 channels, and therefore the Quality of Service of the network.

The default Wi-Fi Rate Adaptation (RA) algorithm used in the Linux kernel depends on the IEEE 802.11 version. Minstrel \cite{minstrel1}\cite{minstrel2} is used for IEEE 802.11a/b/g releases, while Minstrel High Throughput (HT) \cite{minstrelht} is used for IEEE 802.11n. Minstrel HT was developed considering PHY/MAC parameters introduced in IEEE 802.11n. These include the number of spatial streams (1 to 4), channel bandwidth (20/40 MHz) and Guard Interval GI (800/400 ns), which were not considered in the original version of Minstrel. Nevertheless, Minstrel HT performs inefficient random sampling of the environment \cite{smartla}, since it tries to sample every rate of a group -- combination of number of spatial streams, GI and channel bandwidth -- at least once every 100 ms. Minstrel HT reacts with a significant delay \cite{evalmin} in scenarios where the link quality improves, such as when a station (STA) is moving towards the access point (AP). Both problems lead to slower responses in dynamic and fast changing scenarios.

Machine Learning (ML) techniques, including Deep Reinforcement Learning (DRL), have been used to solve network problems for some time and their use is significantly increasing due to recent hardware and software developments. The classical algorithms and heuristic models are designed considering generic scenarios. The wireless channel is highly dynamic; thus, what is optimal for a given scenario may be sub-optimal for another one. With ML techniques, it is possible to autonomously learn the main features of each wireless channel from raw data as well as to develop models that are smarter and self-adaptive in the sense that they adjust to the wireless channel they are facing. In \cite{rw1_35}\cite{rw2_99} the authors provide a comprehensive survey on the latest research efforts in this field. The two most relevant conclusions are: 1) most works use simplistic simulation data that does not reflect the complexity of real world scenarios; 2) most practical implementation of ML models in wireless platforms are not realistic, because the existing computing facilities in network systems are not designed to support the training overhead, due to the need of collecting and transferring large amounts of data. 

The main contribution of this paper is a simple DRL-based Wi-Fi RA algorithm named Data-driven Algorithm for Rate Adaptation (DARA). Taking advantage of DRL techniques, DARA is able to learn the optimal Wi-Fi MCS to be used in a given time interval by calculating the average Signal-to-Noise Ratio (SNR) of the frames received in the previous time interval at the transmitter (TX) node, thus being standard-compliant. DARA is trained and evaluated using the ns-3 simulator \cite{ns3} and \emph{ns3-gym} \cite{ns3gym}, which is a framework that enables the development of an OpenAI Gym RL interface \cite{oaigym} for the ns-3 simulator. Together they allow RL algorithms to be trained and evaluated using an underlying ns-3 simulation as the model environment. DARA's performance is compared with state of the art RA algorithms, namely Minstrel HT and Ideal \cite{ideal}. At first, DARA was evaluated using plain simulation models, considering both static and mobile scenarios. Then, it was evaluated using trace-based simulation models \cite{tracebased1}\cite{tracebased2} that used experimental traces of SNR and the positions of the nodes \cite{simbed}\cite{simbedplus} previously captured from the w-ilab.2 testbed \cite{wilab2}, part of the FED4FIRE+ \cite{fed4fire} project. The trace-based simulation approach provided repeatability and reproducibility in ns-3 of real experiments, using real captured data; besides this advantage, it also allowed a fair comparison between DARA, Minstrel HT and Ideal operating in the same exact conditions.

The rest of the paper is organized as follows. Sections II discusses the Related Work. Section III explains the proposed DARA algorithm and its implementation. Section IV evaluates DARA in simulation. Finally, Section V provides some concluding remarks and points out the future work.

\section{Related Work}

Different solutions have been proposed to solve the Wi-Fi RA problem. We can classify them as classic algorithms, heuristic models or ML models and further divide them into open-loop and closed-loop. The open-loop approaches rely on sampling the environment without any kind of feedback from the receiver node. The closed-loop approaches use a feedback mechanism to obtain information from the receiver node. There are also hybrid approaches \cite{easira} that use both open and closed-loop methods. 

Two examples of open-loop approaches are Minstrel HT and STRALE. \textbf{Minstrel HT} \cite{minstrelht} is the default RA algorithm used in the ath9k \cite{ath9k} wireless driver present in Linux kernel. It is the upgraded version of Minstrel, designed for legacy IEEE 802.11a/b/g devices. Every 100 ms, Minstrel HT uses probing frames to sample all possible rates -- at least once -- for a spatial stream number, channel bandwidth and GI configuration. Then, it keeps a table with statistical information to calculate expected frame loss ratios and throughputs for each combination. Finally, it chooses the top MCS configurations for the transmission of data frames. However, it performs inefficient random sampling of the environment and adapts slowly in some scenarios. \textbf{STRALE} \cite{strale} is an open-loop RA algorithm that adapts the MCS and length of frame aggregation together. STRALE determines the optimal frame aggregation length using an Exponential Weighted Moving Average for each Block ACK received. However, it does not take SNR into consideration.

Two examples of closed-loop approaches are ARAMIS and OFRA. \textbf{ARAMIS} \cite{aramis} adapts -- on a per-packet basis -- the data rate and channel bandwidth in IEEE 802.11n links. The authors observed that Channel State Information (CSI) is still not widely supported across all chipsets. To overcome this difficulty, they devised an easier calculable channel quality metric named \emph{diffSNR}. However, despite being standard-compliant, retrieving feedback embedded in the Acknowledgement (ACK) frames, using the MCS Request and MCS Feedback \cite{80211nsta} mechanism, was not supported by most commodity IEEE 802.11n chipsets. \textbf{OFRA} \cite{ofra} -- On Demand Feedback Rate Adaptation -- is a closed-loop approach that estimates the channel quality based on the Signal and Interference to Noise Ratio value observed in the receiver. The receiver node has a lookup table with a set of thresholds at which data rates should be changed. It transmits that information back using ACK frames. However, OFRA is not compliant with the IEEE 802.11 standard in scenarios with ACK-less traffic wherein it uses a custom-made feedback frame.

In recent years, there have been a few proposals such as \cite{smartla} and \cite{EDrivenACra} that considered DRL to address the link adaptation problem. DRL provides the intelligence and scaling benefits that are important with the increasing degrees of freedom associated with link configuration.

The authors of \cite{smartla} propose an adaptive automated on-line learning mechanism, called ``Smart Link Adaptation" (SmartLA). SmartLA is a reinforcement learning model that learns to dynamically configure link parameters available in the IEEE 802.11n/ac standards -- such as channel bonding, MCS values, SGI and frame aggregation (A-MPDU) -- considering the information extracted from the receiver ACK frames -- such as SNR, Bit Error Ratio and Frame Error Ratio (FER). In \cite{smartla} the performance of SmartLA was evaluated both in simulation and experimentally. The results proved that SmartLA could significantly boost the overall network performance in average throughput, average packet delay and average packet loss ratio, for different scenarios. However, the authors refer that SmartLA still needs to be tested at a large experimental testbed as it remains unclear how the machine learning algorithm scales in real time environments.

We present a simple yet robust link adaptation scheme, based on RL that addresses the drawbacks identified in the described solutions, while being standard-compliant.

\section{Data-driven Algorithm for Rate Adaptation}

In this section, we first overview the RL and DRL concepts. Then, we detail the DARA DRL model and the implementation used for training and evaluation purposes.

\subsection{Reinforcement Learning Overview}

\begin{figure}
  \centering
  \includegraphics[width=.8\linewidth]{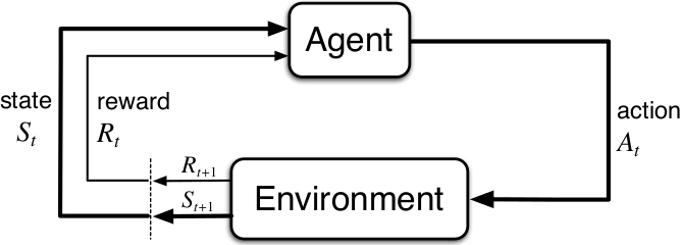}
  \caption{Reinforcement learning model.}
  \label{fig:RLdiag}
\end{figure}

The RL model, described in Figure \ref{fig:RLdiag}, is composed of the agent and the environment. The communication between them is made using signals of state, action and reward. The agent learns the decisions (actions) to make through the observations (states) received from the environment and its decision is evaluated through the reward. The objective of RL is to learn a policy of actions to take for a given environment state that maximizes the overall cumulative reward for every state/action pair. There are multiple learning algorithms available in the literature. Q-learning \cite{qlearning} is a model-free algorithm, which means that it does not use any prior knowledge of the environment but learns from trial and error. Q-learning only works with a discrete action space $A$. The objective of Q-learning is to learn the optimal policy that maximizes the expected cumulative reward. Q-learning learns and updates its Q-function values via trial and error using \cref{eq:qlearning}, where $Q(s,a)$ is the expected cumulative reward when the agent selects action $a$ in state $s$ considering that future actions are selected according to the learnt policy. $r(s,a)$ is the reward for taking the action $a$ in state $s$, and $\max_{a \in A}Q(s_{new},a_{all})$ is the maximum possible reward of the new state, result of the current action, where $s_{new}$ is the new state and $a_{all}$ represents every $a \in A$. The learning rate $\alpha$ determines the rate at which new values update the total Q-value. Finally, the discount factor $\gamma \in [0,1]$ determines the importance of future rewards in the calculation of the expected cumulative reward.

\begin{equation}
Q(s,a)\leftarrow (1-\alpha)Q(s,a)+\alpha[r(s,a)+\gamma\max_{a \in A}Q(s_{new},a_{all})]    
\label{eq:qlearning}
\end{equation}

Q-learning is not designed to cope with scenarios where there is a scaling amount of possible observations and actions. For that reason Deep Q-Networks (DQN) \cite{dqn} have been proposed. Instead of having an infinitely large Q-table, DRL uses a Deep Neural Network with parameters $\theta$ as a function approximator of the optimal Q-value (i.e., $Q^*(s,a) \approx Q(s,a;\theta)$.

\subsection{DARA DRL model}

In this work, we consider single dimension states and actions. The objective is to train an agent that learns the optimal MCS that should be used for a fixed time interval $t_n$, based on the SNR of the received frames from the previous interval $t_{n-1}$. In DARA we define the action space $A$ as a subset list of the MCS values available in the IEEE 802.11n standard. We consider scenarios with Single Input Single Output (SISO) channels with fixed bandwidth and GI. Thus, eight possible \textbf{actions} ($MCS_{0}$ to $MCS_{7}$) are considered. The \textbf{state} is defined as the average SNR value observed during the time interval $t_{n-1}$, considering each frame arriving at the TX node. Finally, the \textbf{reward} is defined in Eq. \ref{eq:reward}. It is the product of the frame success ratio and the normalized MCS of the transmitted frames during the time interval $t_{n-1}$.

\begin{equation} \label{eq:reward}
reward=\frac{MCS_{n}}{MCS_{7}} \times FER, \: n \in [0,1,...,7]
\end{equation}

With the definition of this reward, we intend to value the combination of success ratio and the highest possible MCS. This avoids the creation of a policy where the agent would consistently pick: 1) the lowest MCS to guarantee the highest frame delivery ratio, regardless of the link condition; or 2) the highest MCS, even if the frames were not being delivered. 

\subsection{DARA Implementation}

\begin{figure}
  \centering
  \includegraphics[width=\linewidth]{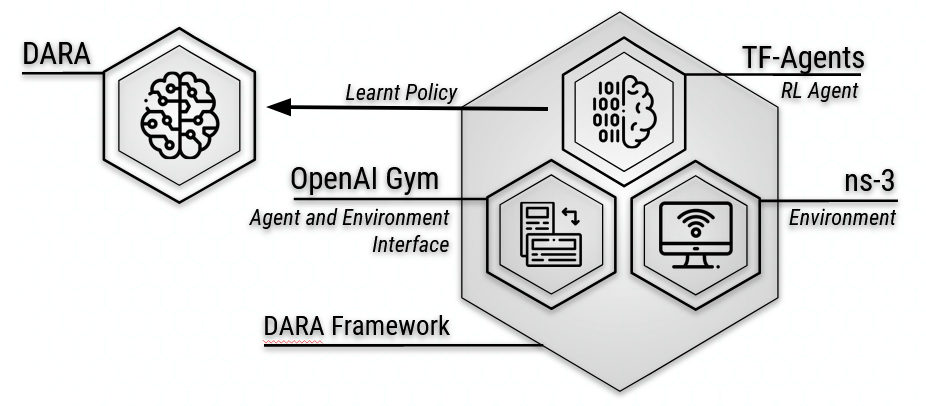}
  \caption{Illustration of the DARA implementation.}
  \label{fig:framework}
\end{figure}

Figure \ref{fig:framework} shows the framework and tools used in the implementation of DARA. We used ns-3 together with the \emph{ns3-gym} framework \cite{ns3gym} to develop the environment, and TF-Agents \cite{TFAgents} as the library to implement the RL Agent. In this section, we go through the ns-3 implementation and explain the actions, observations and reward.

\subsubsection{ns-3 Environment}

We use ns-3 (version 3.35) and the \emph{ns3-gym} framework to train DARA in a wide variety of scenarios. Table \ref{ns3config} summarizes the ns-3 configurations used. 
The traffic generation saturates the link with packets being permanently sent from the TX node to the receiver (RX) node, throughout the entire duration of the simulation.

\begin{table}
\centering
\caption{ns-3 Environment Configuration Parameters}
\begin{tabular}{ll}
\hline
\textbf{Configuration Parameter} & \textbf{Value}                                      \\ \hline
Wi-Fi Standard          & IEEE 802.11n                               \\
Propagation Delay Model & Constant Speed                             \\
Propagation Loss Model  & Friis                                      \\
Frequency               & 5180 MHz                                   \\
Channel Bandwidth       & 20 MHz                                     \\
Wi-Fi MAC               & Ad-hoc                                     \\
Remote Station Manager  & DARA (Constant Rate) / Ideal / Minstrel-HT \\
Traffic                 & UDP                                        \\
Simulation Statistics   & FlowMonitor \cite{flowmon}                               \\ \hline
\end{tabular}
\label{ns3config}
\end{table}

\subsubsection{Action}

In its current version, DARA only supports one MCS per time interval $t_n$, corresponding to the MCS level for long GI, 20 MHz channel bandwidth and SISO IEEE 802.11n operation. At each time interval $t_n$ the new MCS that comes from the agent is selected.

\subsubsection{Observation}

The number of received frames in the TX node and their SNR is tracked. Then, the average SNR for the time interval $t_n$ is calculated. When the agent selects an MCS higher than what the receiver can decode due to poor link conditions, the frames will not be successfully delivered. In such scenario, there is no observation to feed the agent, because the SNR is calculated based on the received frames. To address this problem, we assume that the SNR of that time interval is zero so that the Agent picks the lowest MCS in the following interval. The implementation of a ``backup rate" to reduce the impact of the problem in the overall performance of the solution is left for future work.

\subsubsection{Reward}

the reward calculation depends on whether the transmitted frame was ACKed or not. From the TX node, there are three possible scenarios for a frame: 1) The frame is successfully received and the TX node knows that information from the corresponding ACK/Block ACK; 2) The frame is not received successfully and the TX node knows that information from the received Block ACK; 3) Neither the ACK/Block ACK nor the Block ACK request are delivered successfully during the timeout period. Then, the FER is calculated based on these occurrences and the MCS selected based on the action taken for that interval.

\subsubsection{Agent Architecture}

The DARA agent is implemented in Python using the TF-Agents library \cite{TFAgents}. From the perspective of the agent, it needs to receive from the environment the \emph{step} that consists of an observation and reward of the previous decision interval, and provide the \emph{action} for the next period. The agent was implemented using the Deep Q-Network (DQN) learning algorithm. In what follows, we detail the main parameters of the DQN learning algorithm:
    \begin{itemize}
        \item \textbf{Observation space} -- one-dimensional float ranging between 0.0 and 1.0. We scaled the value of the SNR by dividing the final SNR value in dB by 100, as unscaled input variables can result in a slow or unstable learning process.
        \item \textbf{Action space} -- one-dimensional integer value, representing the MCS used to transmit the following frames.
        \item \textbf{Optimizer} -- Adam \cite{AdamOpt} with a learning rate of $10^{-2}$ that scales the magnitude of the weight updates to minimize the network loss function. As for the loss function, we used the standard mean square error function.
        \item \textbf{Epsilon greedy} -- a value between 0 and 1 that determines the percentage of actions taken by the agent in an exploratory way rather than an exploiting way. This value starts at 1.0 (every decision is random) and, through a polynomial decay, decreases over a defined period until 0.1 (10\% of the decisions are random).
        \item \textbf{Q-Network} -- 2 layers of fully connected parameters with 32 units each. The target Q-Network has the same structure.
        \item \textbf{Replay Buffer} -- with a size of $10^{6}$ trajectories. As the simulation progresses, trajectories are appended to the replay buffer. Each time the agent is trained, it randomly samples a batch of 64 trajectories from the replay buffer.
    \end{itemize}

The agent runs using two types of sessions:

\begin{itemize}
    \item \textbf{Training session} -- The training process starts by filling the replay buffer with the trajectories collected from the simulation. A trajectory consists of a time step (i.e., the initial observation of the environment), the action step (i.e., the action that was taken considering the previous time step), and the next time step (i.e., the new observation and the reward that was obtained using the previous action step). This replay buffer is filled while the simulation is running until the \textbf{Game Over Module} reports the episode is over. The agent is then trained randomly grabbing from the replay buffer an amount of trajectories defined with the hyper parameter \emph{batch size} updating the weights accordingly and increasing the training step counter. The user can then adapt the number of episodes the simulation runs, how frequent the training happens and the way epsilon greedy adjusts over the training process. When the training is over or we want to pause it, we can save the progress with a checkpoint so that the current state of the policy can be recovered later on.
    
    \item \textbf{Evaluation session} -- We load the trained policy and assume a fixed epsilon greedy factor of zero to avoid exploratory attempts for DARA. When doing evaluation sessions we initialize the simulation in evaluation mode to log the throughput obtained at the application level (and packet capture files, if necessary, for debug purposes). In addition, we keep track of the rewards throughout the session and log the average reward of the episode.
\end{itemize}

\section{Simulation Results}

In this section we go through the preliminary validation and the simulation scenarios used to evaluate DARA. The rationale was to start first the validation of DARA in plain simulation scenarios before moving to more realistic scenarios using trace-based simulations. Finally, we discuss the results.

\subsection{Preliminary Validation}

The following results allow comparing the performance of DARA with Minstrel HT and Ideal RA algorithms. The Ideal RA algorithm available in ns-3 \cite{ideal} maintains in every station the SNR of every packet received and sends back this SNR to the sender by an out-of-band mechanism. Each sender keeps track of the last SNR sent back by a receiver and uses it to pick an MCS based on a set of SNR thresholds derived from a target Bit Error Ratio (BER) and MCS-specific SNR/BER curves. From now on, we refer to Minstrel HT and Ideal as \textbf{MIN} and \textbf{ID}, respectively.

To validate whether the logic behind the model was well defined we started with a simple scenario with two fixed nodes at a distance of 5 meters. As expected, the optimal MCS for these conditions was MCS 7 and \textbf{DARA} learnt accordingly, thus preliminarily validating its correct operation. The next step was to teach DARA what SNR value ranges should be considered for each of its actions. The objective was to replicate the behaviour of the ID algorithm. To achieve this, we used a scenario where the RX node steadily moved away from the TX node until their communication link was lost.

\begin{figure}
  \centering
  \includegraphics[width=\linewidth]{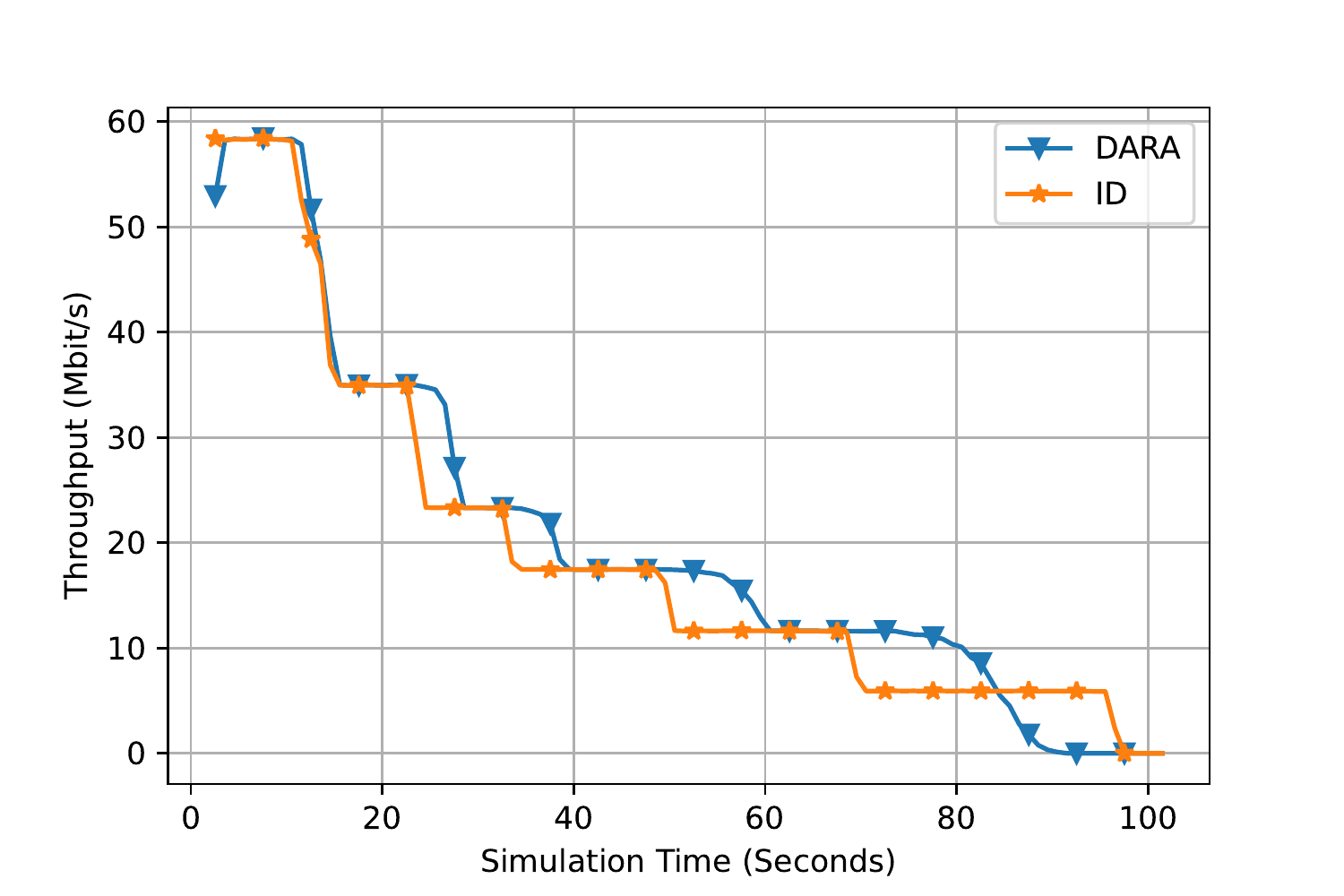}
  \caption{Performance comparison between ID and the trained version of DARA.}
  \label{fig:train}
\end{figure}

In Figure \ref{fig:train} we plot the UDP throughput at the application level throughout the simulation of the training scenario. Here we can observe the well-defined SNR thresholds of the ID algorithm with each ``step" representing a different MCS. From the results we concluded that DARA adjusted the SNR thresholds differently, achieving higher average throughput when compared to ID.

\subsection{Plain Simulation Scenario}

In order to check how DARA compares with MIN and ID when the radio channel quality changes randomly, we have considered a scenario where every two seconds a new random position, between zero and 600 meters, is assigned to the RX node. The simulations in this scenario were run, using the IEEE 802.11n standard in the 5 GHz frequency band. A channel bandwidth of 20 MHz and transmitting power of 20 dBm were configured in each node, with no antenna gains. Constant speed was used for the propagation delay, Friis for the propagation loss and NIST for the error rate model. UDP traffic was permanently generated above link capacity to keep the link saturated. The TX node was at a fixed position and the RX node had its position changed dynamically to simulate the random variation of the SNR.

\begin{figure}
  \centering
  \includegraphics[width=\linewidth]{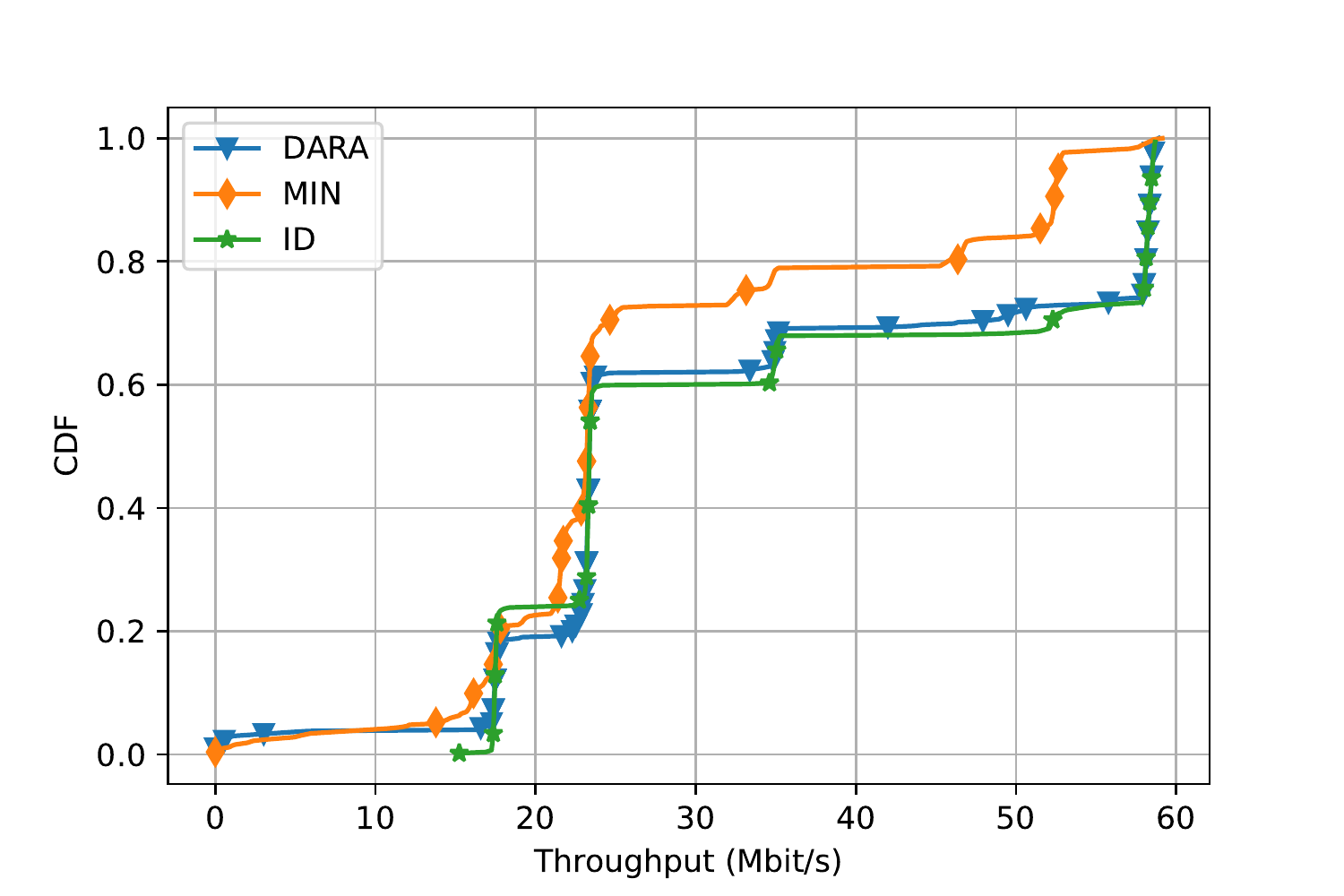}
  \caption{Cumulative distribution function (CDF) of UDP throughput when the radio channel quality changes randomly.}
  \label{fig:cdfsim}
\end{figure}

Figure \ref{fig:cdfsim} shows the throughput obtained for each RA algorithm. The average throughput for DARA and ID was 32.8 Mbit/s and 33.8 Mbit/s, respectively, while for MIN we obtained 27.6 Mbit/s. 

From these results and the plot of Figure \ref{fig:cdfsim}, we can conclude that, when the radio channel quality changes randomly, DARA is able to reach the performance of ID and surpass MIN.

\subsection{Trace-based Simulation Scenarios}

\begin{figure}
  \centering
  \includegraphics[width=\linewidth]{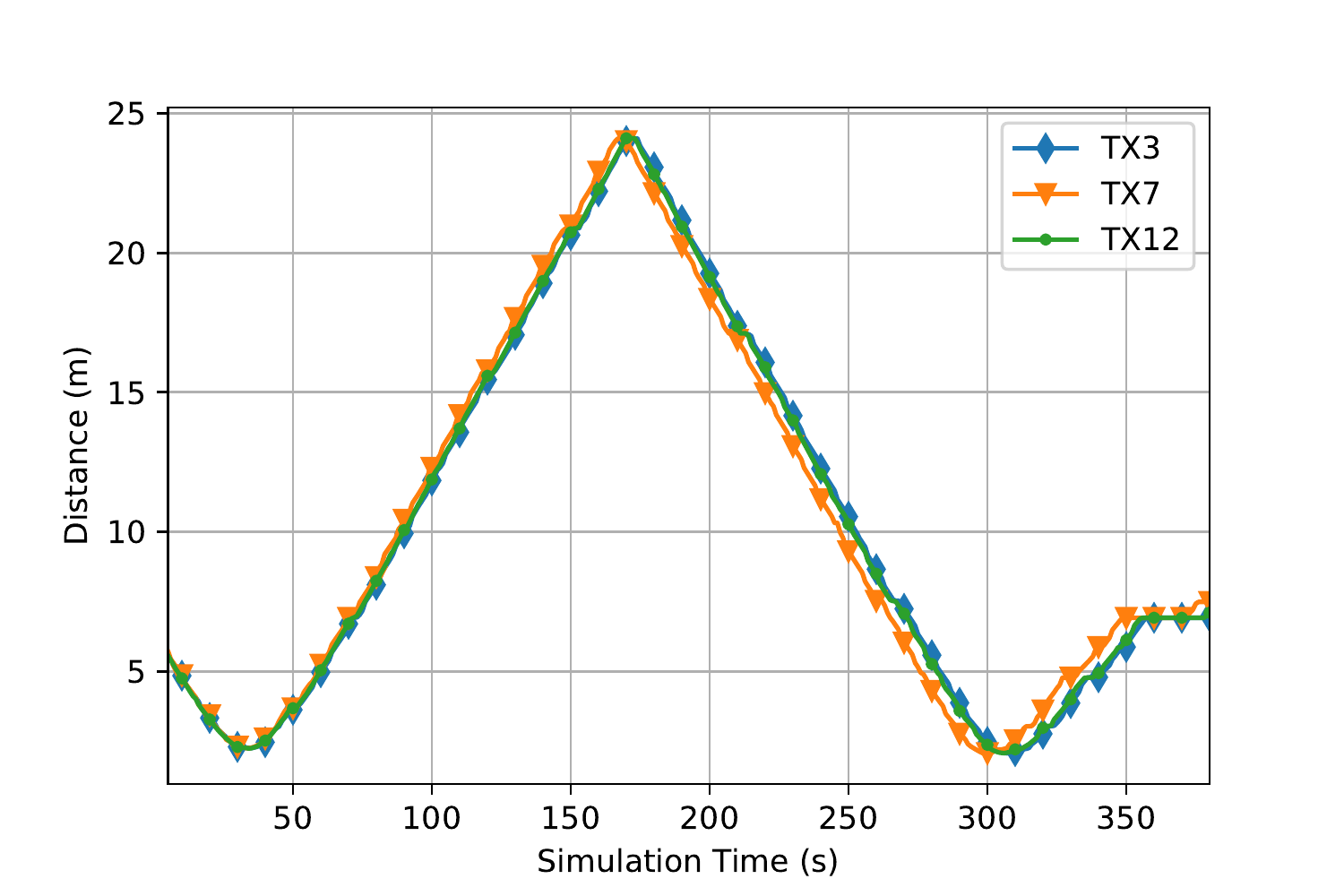}
  \caption{Variation of the distance between the TX node and the mobile node in three trace-based scenarios.}
  \label{fig:3traceScenarios}
\end{figure}

The trace-based simulation scenarios used the same simulation configuration of the plain simulation scenario. When using trace-based simulations, we bypass every \emph{ns-3} class that would have an impact in the SNR calculation of a packet (e.g., the distance between nodes or the propagation delay model). Instead, we force the SNR to values previously captured in a real experiment with two nodes from the w-ilab.2 testbed \cite{wilab2}. The real experiments considered a fixed TX node and a mobile RX node, operating in SISO mode, using a frequency of 5220 MHz and 20 MHz bandwidth. Both antennas had gains of -7 dBi and each node was configured with the same TX power. In this work we are considering three scenarios with different TX powers: 12 dBm, 7 dBm and 3 dBm. From now on, we will refer to each trace-based scenario as \textbf{TX12}, \textbf{TX7} and \textbf{TX3}. The variation of the distance between the TX node and the mobile node for each scenario is represented in Figure \ref{fig:3traceScenarios}. 

\begin{figure}
  \centering
  \includegraphics[width=\linewidth]{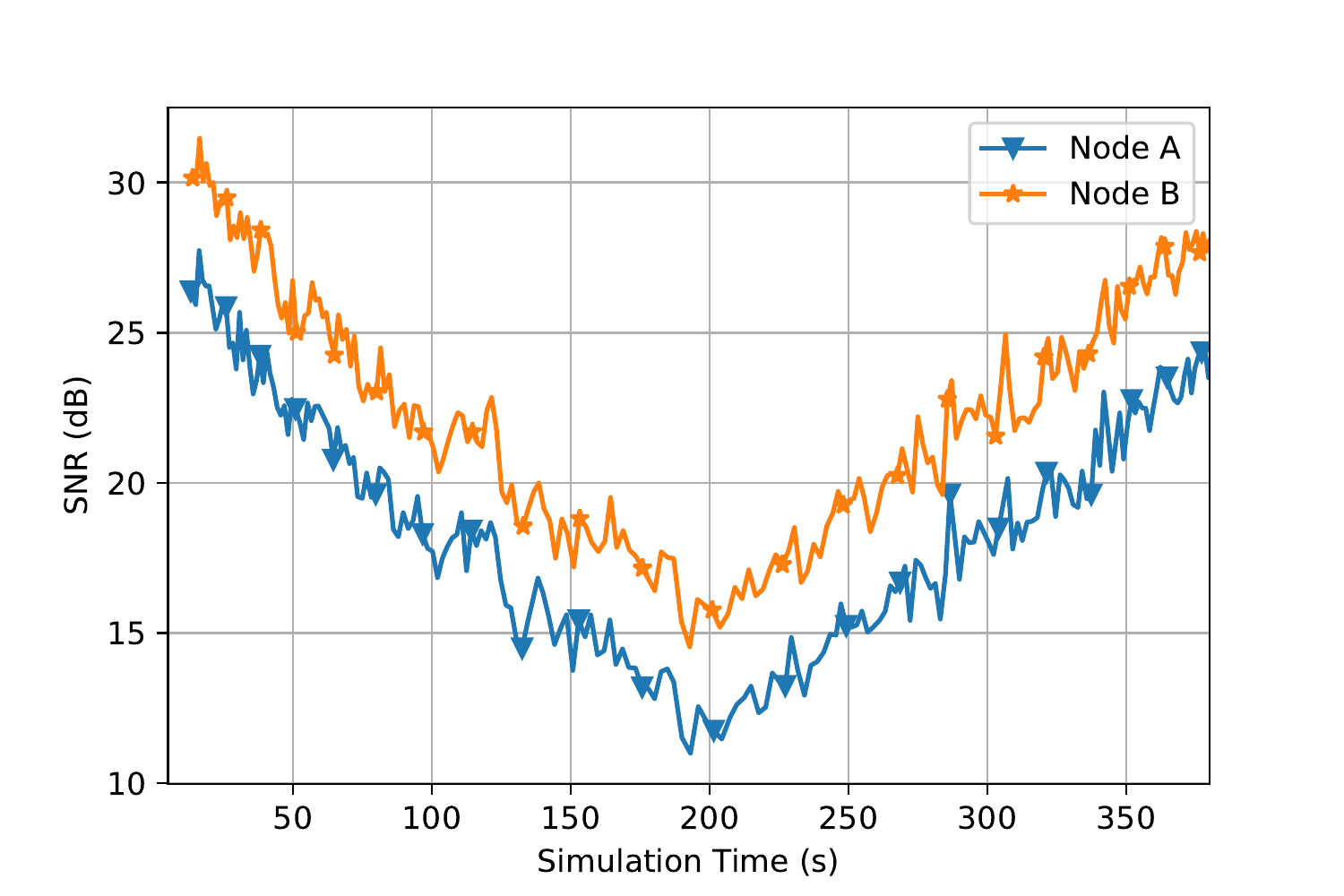}
  \caption{Asymmetry between the two nodes in the TX12 experimental scenario.}
  \label{fig:12snrgap}
\end{figure}

Despite having identical hardware and being configured in the same way, in practice the radio link was asymmetric. Figure \ref{fig:12snrgap} presents the asymmetry for the TX12 scenario. In that sense, it was necessary to adjust DARA's policy for the asymmetry of each specific scenario. We did that with a single training session, from scratch, before each evaluation session. 

\begin{figure}
    \centering
    \subfloat[A $\rightarrow$ B.]{
        \includegraphics[width=\linewidth]{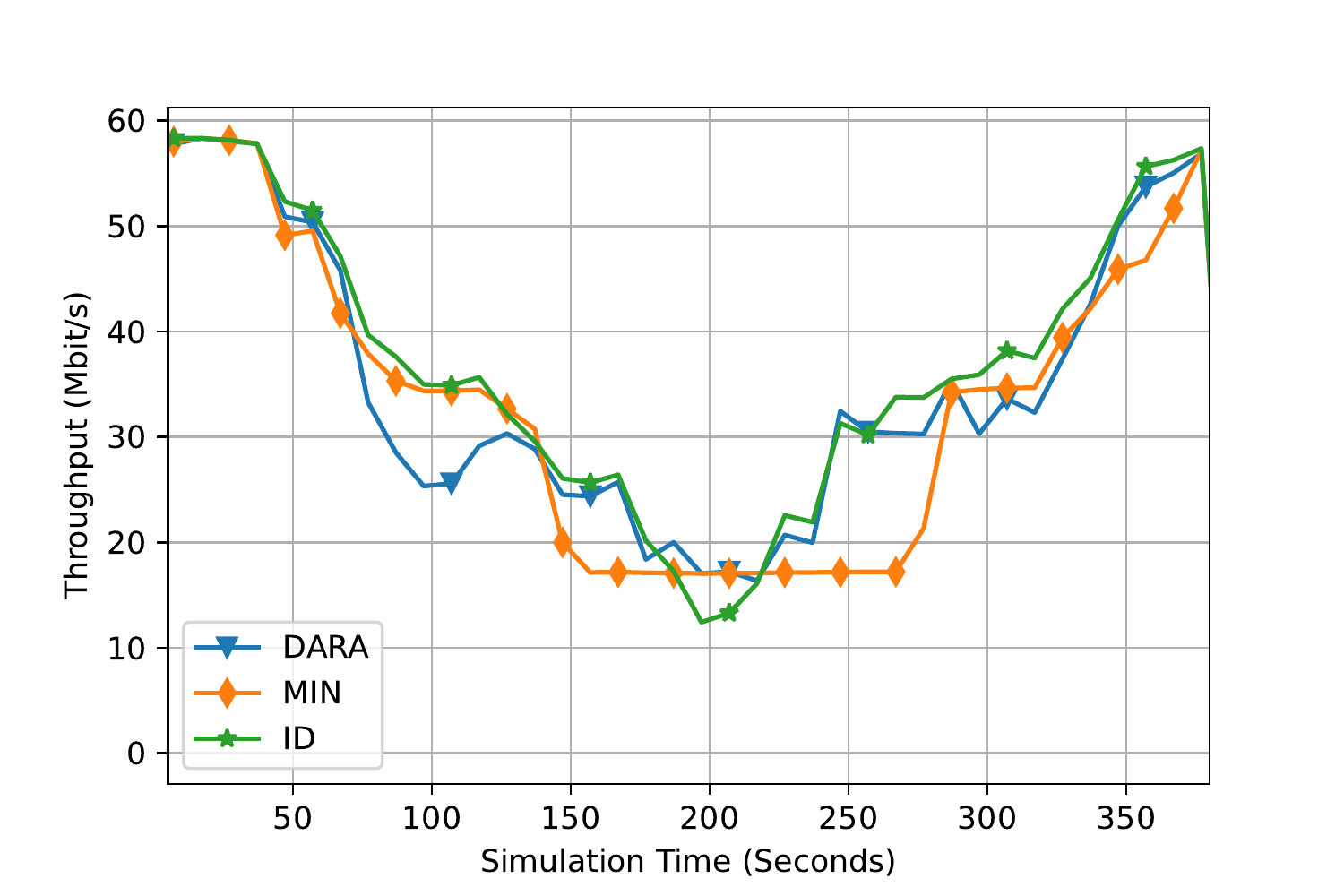}
    }
    \vfil
    \subfloat[B $\rightarrow$ A.]{
        \includegraphics[width=\linewidth]{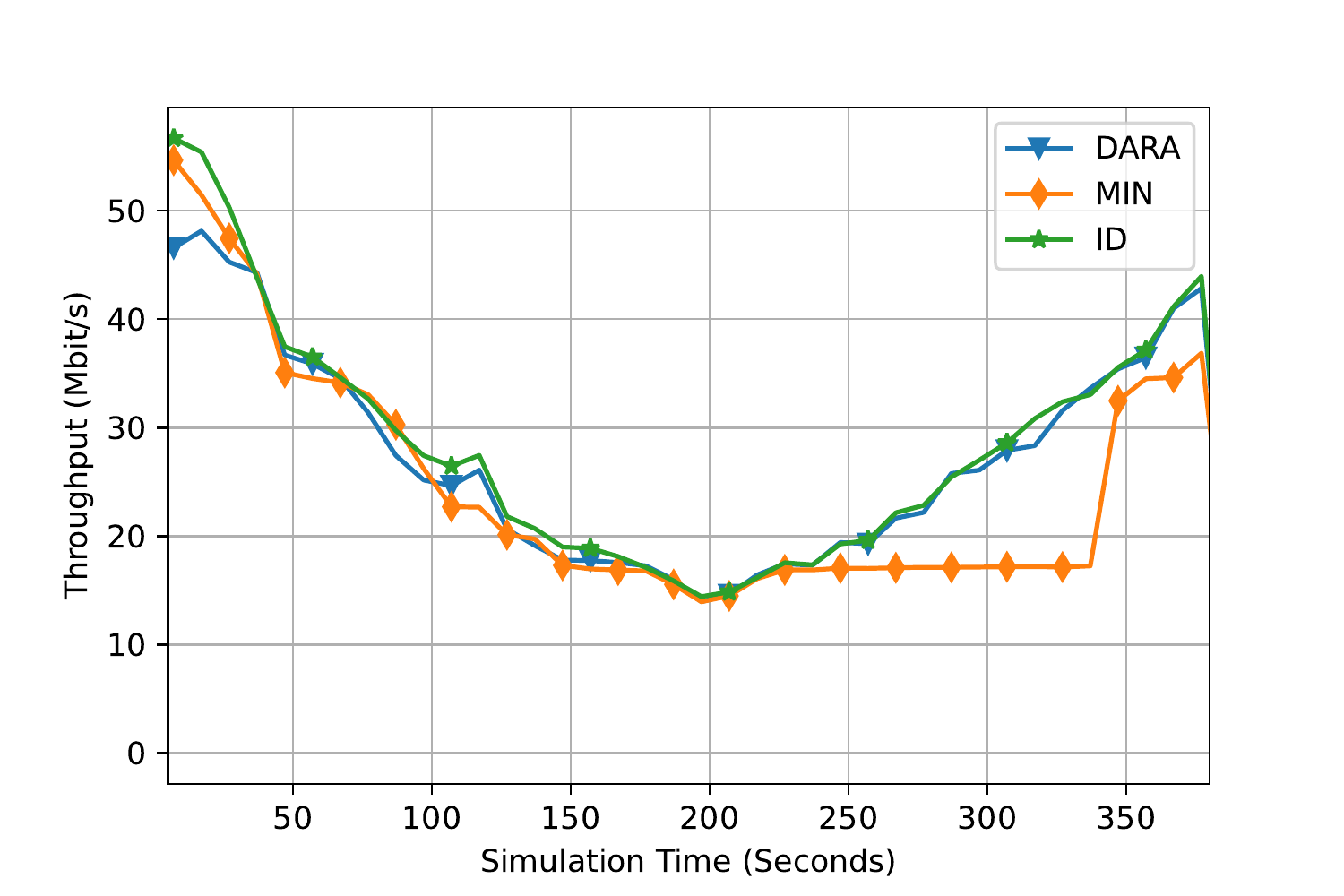}
    }
    \caption{Throughput over simulation time for the TX12 scenario.}
    \label{fig:12dbmth}
\end{figure}

For each scenario, we saturated the link with UDP traffic. Traffic was generated in both directions: A $\rightarrow$ B and B $\rightarrow$ A. As an example of the channel asymmetry, we considered the TX12 experimental scenario as shown in Figure \ref{fig:12snrgap}. The average throughput over the simulation period in both directions is shown in Figure \ref{fig:12dbmth}. DARA was able to learn the channel asymmetry solely using the SNR observed from the TX node side. MIN adjusts better when the link quality degrades. When the mobile node is moving back towards the TX node, the link quality improves. However, in these situations, the reaction of MIN is significantly slower when compared with DARA and ID. This behaviour happens in every evaluated scenario.

\begin{figure*}
    \subfloat[TX3, A $\rightarrow$ B.] {
        \includegraphics[width=0.3\linewidth]{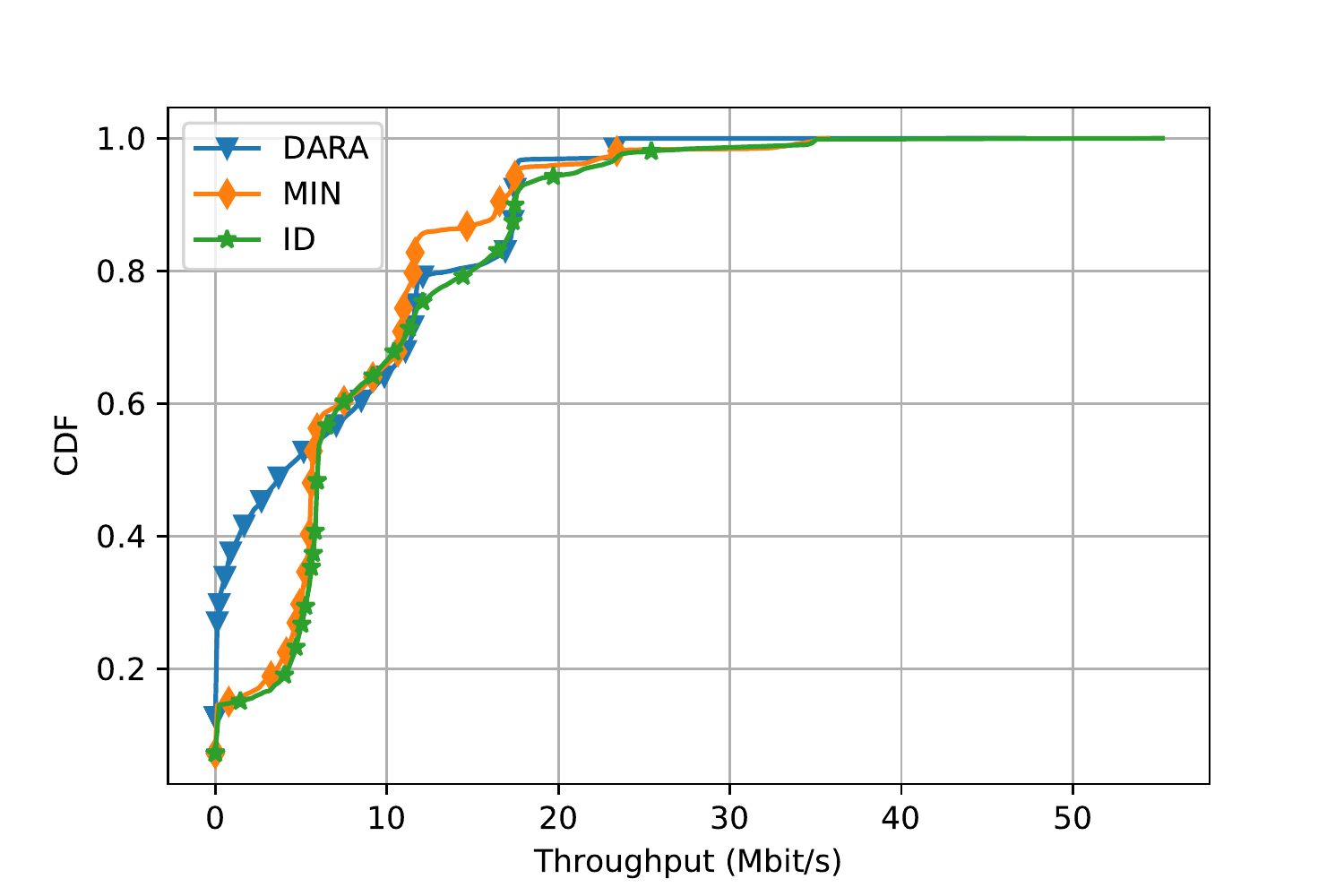}
        \label{fig:3cdfnormal}
    }
    \hfil
    \subfloat[TX7, A $\rightarrow$ B.] {
        \includegraphics[width=0.3\linewidth]{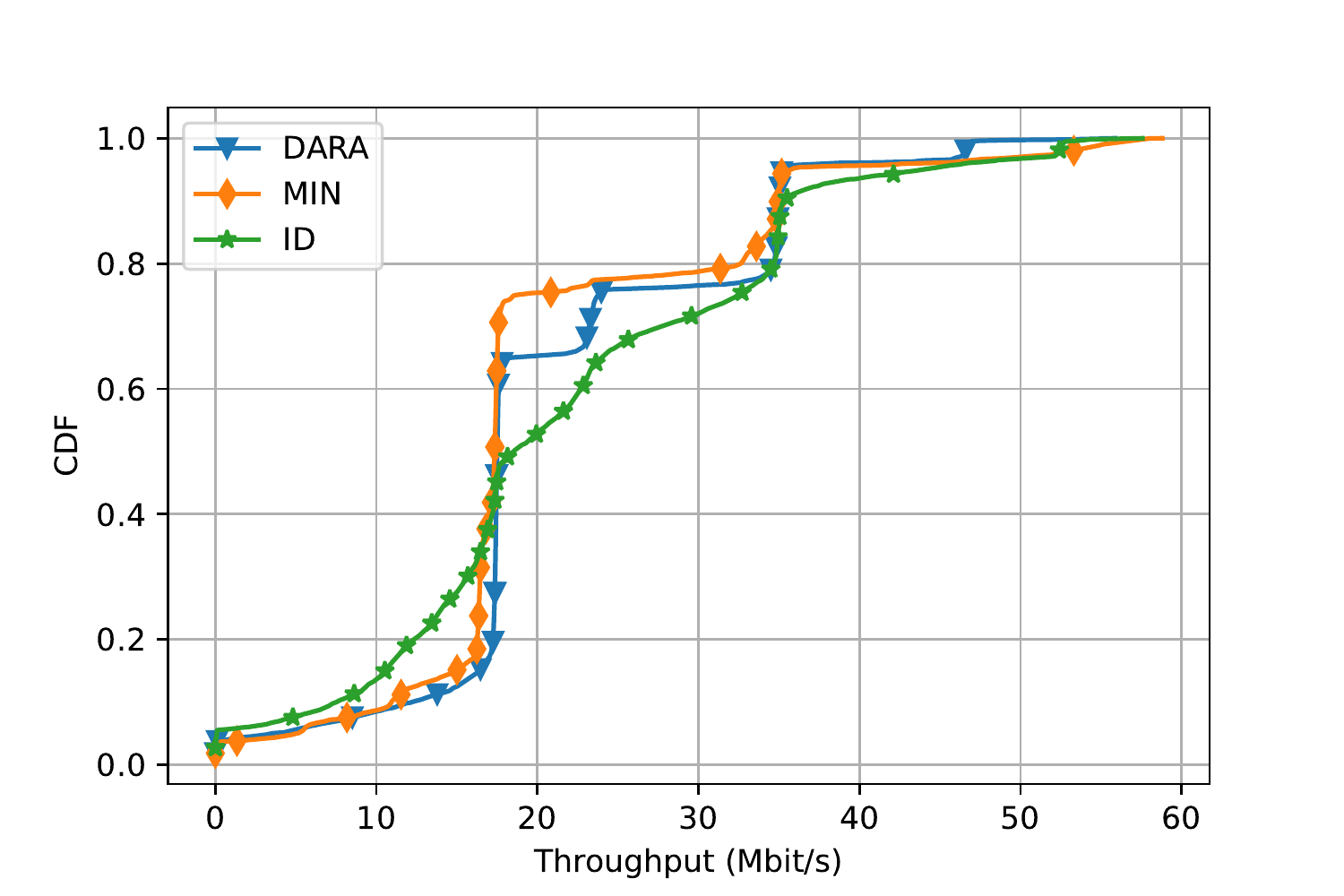}
        \label{fig:7cdfnormal}
    }
    \hfil
    \subfloat[TX12, A $\rightarrow$ B.] {
        \includegraphics[width=0.3\linewidth]{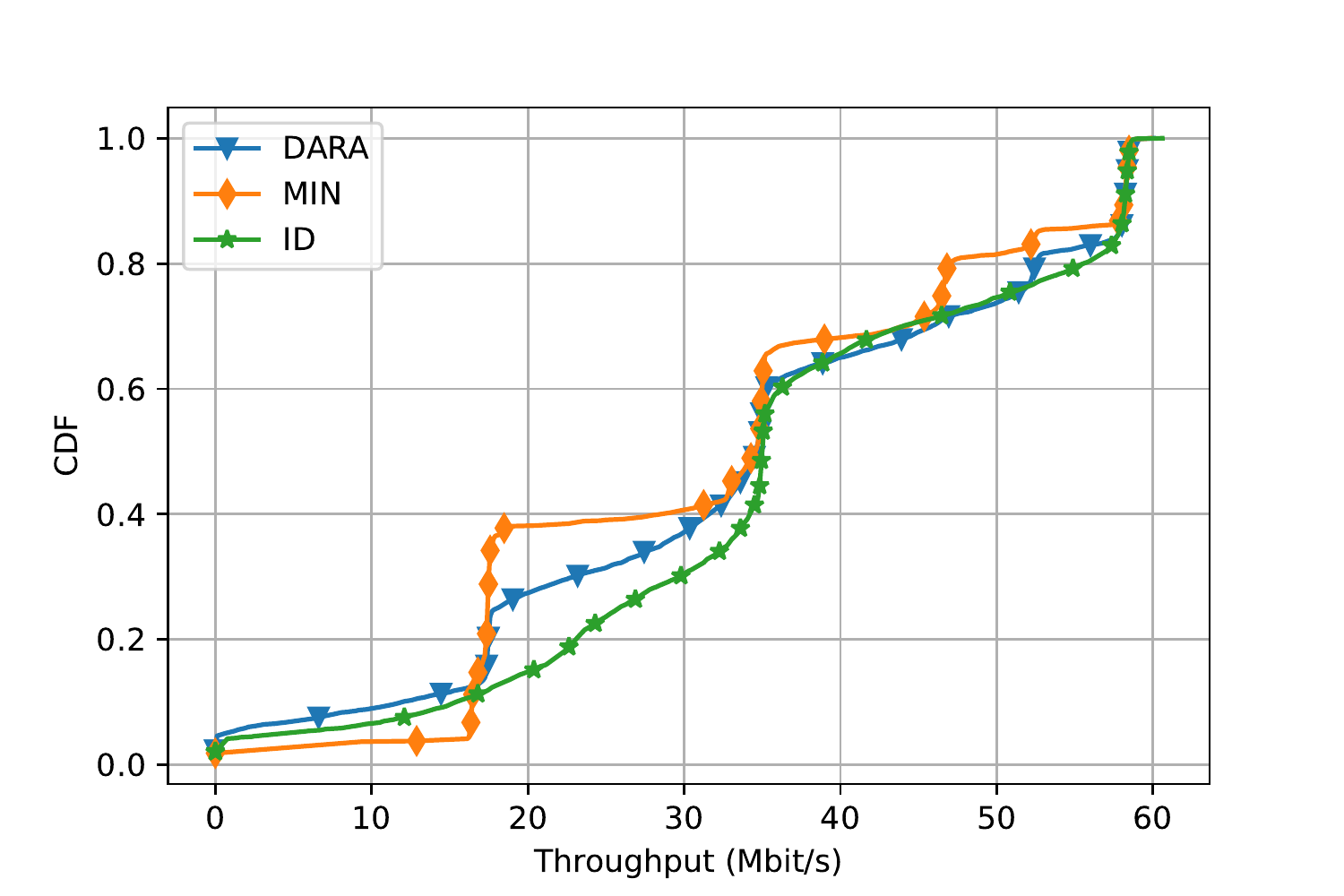}
        \label{fig:12cdfnormal}
    }
    \vfil
    \subfloat[TX3, B $\rightarrow$ A.] {
        \includegraphics[width=0.3\linewidth]{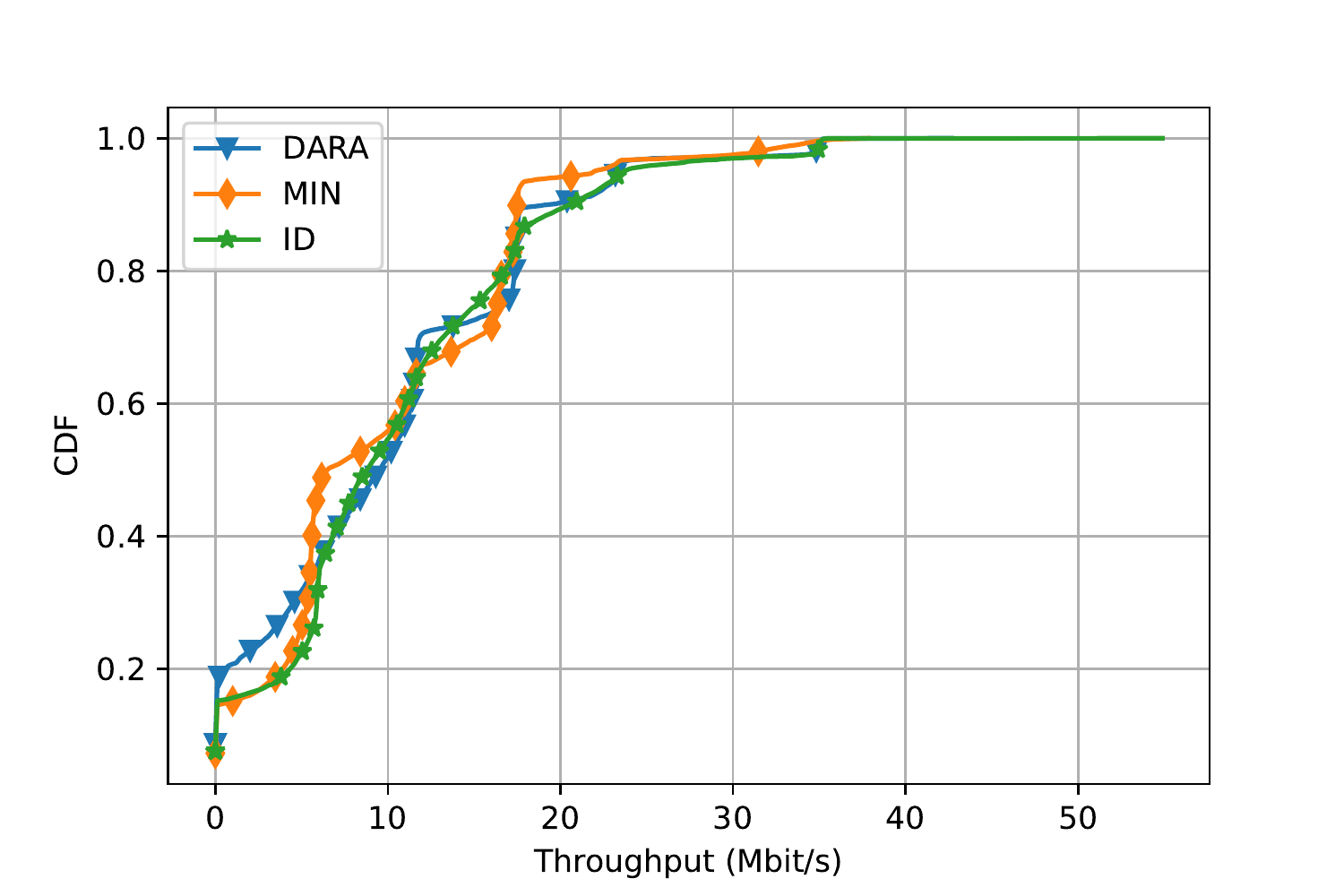}
        \label{fig:3cdfreversed}
    }
    \hfil
    \subfloat[TX7, B $\rightarrow$ A.] {
        \includegraphics[width=0.3\linewidth]{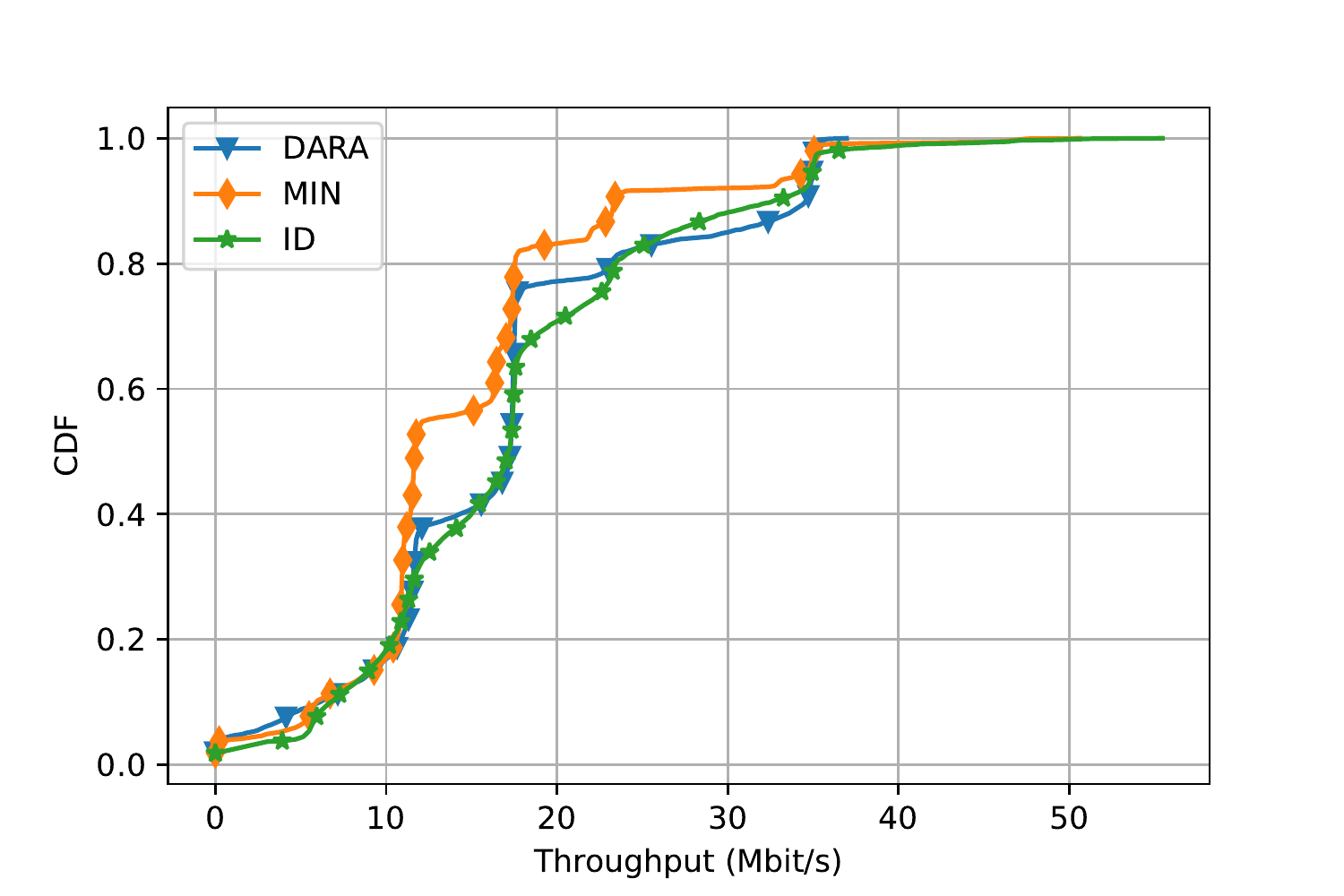}
        \label{fig:7cdfreversed}
    }
    \hfil
    \subfloat[TX12, B $\rightarrow$ A.] {
        \includegraphics[width=0.3\linewidth]{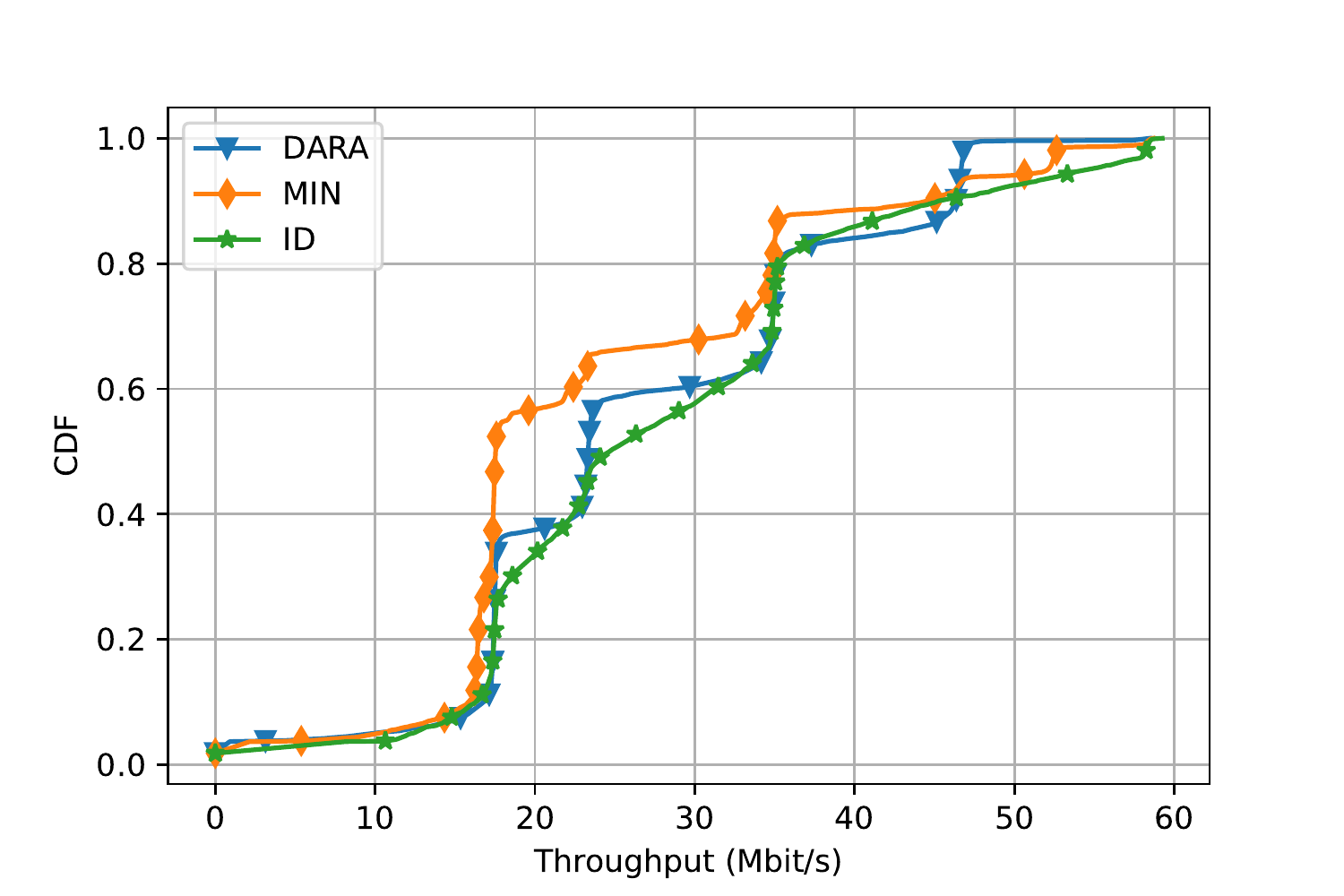}
        \label{fig:12cdfreversed}
    }
    \caption{Simulation CDF Results}
\end{figure*}

Figures \labelcref{fig:3cdfnormal,fig:3cdfreversed,fig:7cdfnormal,fig:7cdfreversed,fig:12cdfnormal,fig:12cdfreversed,} present the Cumulative Distribution Functions (CDF) for the three trace-based scenarios. Figure \ref{fig:3cdfnormal} shows DARA's worst performance. The 40th percentile shows a difference of 4 Mbit/s between DARA and the other RA algorithms, but it improves after the 50th percentile and approximately fitting the other CDFs. Figure \ref{fig:3cdfreversed} shows three balanced CDFs, with no significant differences between RA algorithms. Figure \ref{fig:7cdfnormal} shows at the 20th percentile that both MIN and DARA performed better than ID. However, this changes right after at the 75th percentile, with MIN DARA and ID achieving 19, 24 and 32 Mbit/s, respectively. Figure \ref{fig:7cdfreversed} is the scenario where DARA held better results. At the 90th percentile we see a difference of 11 Mbit/s when compared with MIN and approximately no difference from ID. Finally, Figure \ref{fig:12cdfnormal} and Figure \ref{fig:12cdfreversed} present two scenarios where MIN also had the worst performance and DARA had similar performance when compared to ID.

Table \ref{tabsummary} summarizes the average throughput of every RA algorithm for each evaluated scenario. DARA's worst performance was in the TX3 with A $\rightarrow$ B traffic orientation scenario, having 19.2\% and 11.3\% lower average throughput than ID and MIN, respectively. However, DARA performed better than MIN in every other evaluated scenario achieving up to 14.9\% higher throughput.

\begin{table*}
\caption{Throughput summary in Mbit/s of 3 trace-based scenarios using a saturated link UDP traffic generation.}
\label{tabsummary}
\centering
\begin{tabular}{c|clclcl|clclcl|l}
\cline{2-13}
                           & \multicolumn{6}{c|}{A $\rightarrow$ B}   & \multicolumn{6}{c|}{B $\rightarrow$ A}  &  \\
 &
  \multicolumn{2}{c}{TX3} &
  \multicolumn{2}{c}{TX7} &
  \multicolumn{2}{c|}{TX12} &
  \multicolumn{2}{c}{TX3} &
  \multicolumn{2}{c}{TX7} &
  \multicolumn{2}{c|}{TX12} &
   \\ \cline{1-13}
\multicolumn{1}{|c|}{DARA} & \marktopleft{c1}7.3  & \% diff & 22.45 & \% diff & 36.04 & \% diff & 10.44 & \% diff & \marktopleft{c2}17.9  & \% diff & 28.16 & \% diff &  \\ \cline{1-13}
\multicolumn{1}{|c|}{MIN}  & 8.23 & $\downarrow$ 11.3\%  & 21.41 & $\uparrow$ 4.9\%  & 34.6  & $\uparrow$ 4.2\% & 10.2  & $\uparrow$ 2.4\%  & 15.58 & $\uparrow$ 14.9\% & 25.61 & $\uparrow$ 10\% &  \\ \cline{1-13}
\multicolumn{1}{|c|}{ID}   & 9.04 & $\downarrow$ 19.2\% \markbottomright{c1}  & 23.01 & $\downarrow$ 2.4\% & 38.12 & $\downarrow$ 5.5\% & 10.85 & $\downarrow$ 3.8\% & 18.4  & $\downarrow$2.7\% \markbottomright{c2}& 29.38 & $\downarrow$ 4.2\%  &  \\ \cline{1-13}
\end{tabular}
\end{table*}

\section{Conclusions and Future Work}

With this work, we have shown that there is room to develop a simple data-driven algorithm using a DRL approach. DARA achieved a similar performance when compared to its state of the art counterparts, while adopting a simpler approach that does not require environment sampling and is standard-compliant. Using trace-based simulations that reflect the complexity of real world experiments, we have detected situations where traditional RA algorithms such as Minstrel HT, adjust slowly. Nonetheless, DARA performed similarly to the Ideal RA algorithm in such scenarios.

As future work, we will consider online training methods, for ensuring longevity and relevance of the model in its deployment phase. In addition, we plan to evolve DARA to support the most recent standards such as IEEE 802.11ac/ax

\bibliographystyle{IEEEtran}
\bibliography{refs}

\end{document}